 \documentstyle[preprint,aps,prb]{revtex}
 \begin{document}
 \draft
 \title{Superconductivity and spin triplet collective
 mode in the $t-J$ model close to antiferromagnetic instability}
 \author{Damian Marinaro and Oleg Sushkov}
 \address{School of Physics, University of New South Wales, 
 Sydney 2052, Australia}
 
 \maketitle
 
 \begin{abstract}
 To investigate relations between long-range antiferromagnetic (AF) order,
 superconductivity and two particle triplet collective excitations we consider 
 a  modified two dimensional $t-J$ model at doping close to half filling. 
 The model includes additional hopping 
 $t^{\prime \prime}$ and nearest sites Coulomb repulsion $V$.
 The additional parameters allow us to control closeness of the system
 to the AF instability. We demonstrate the possibility of co-existence of
 long-range AF order and d-g-wave superconductivity.
 In the phase with long-range AF order we find,
 analytically, superconducting gaps and spin wave renormalization.
 We demonstrate 
 that at approaching the point of the AF 
 instability the spin triplet collective excitation
 arises with energy below the superconducting gap.
 \end{abstract}
 

 It is widely accepted now that superconductivity of cuprates is
 closely related to their unusual magnetic properties, and
 it is increasingly clear that magnetic pairing is the most realistic
 mechanism of cuprate superconductivity.
 However the mechanism of pairing as well as other unusual properties
 are far from completely understood.
 The problem has been attacked along several directions.
 First we have to mention the empirical or semi-empirical approach
 which allows one to relate different characteristics measured
 experimentally. This approach is to a large extent based on the
 Hubbard model. For a review see article \cite{Chub}.
 In the low energy limit, the Hubbard model can be reduced to the
 $t-J$ model. Another approach to cuprates is based on numerical
 studies of the $t-J$ model (see review \cite{Dag}).
 Our studies are also based on this model. We used the
 ordered Neel state at zero doping as a starting point to
 develop spin-wave theory of pairing \cite{Flam}. The method we used was 
 not fully satisfactory, since it violated spin-rotational symmetry,
 nevertheless it allowed us to calculate from first principles all the most
 important properties including critical temperature, spin-wave pseudogap 
 and low energy spin triplet excitations \cite{Sus}. 
 
 A sharp collective mode with very low energy has been revealed in YBCO in
 spin polarized inelastic neutron scattering \cite{Ros,Dai,Fong}.
 A number of theoretical explanations have been suggested for this 
 effect \cite{Dem,Sus}. All of those explanations are based on the idea that
 the system is close to AF instability. However, all known
 explanations use some uncontrolled approximations and assumptions.
 
 In the present work we investigate a close to half filling regime for the  
 2D $t-J$ model where it 
 can be solved analytically without any uncontrolled approximations.
 It can be done for the region of parameters where long-range AF order is 
 preserved under doping. We analyze the superconducting pairing in this regime
 and consider spin triplet collective excitation. It is demonstrated
 that the narrow collective mode arises at approaching the point of AF
 instability and the energy of the mode is below the superconducting gap. 
 
 Let us consider a $t-J-J^{\prime \prime}-V$ model defined by the
 Hamiltonian
 \begin{equation}
 \label{H}
 H=-t\sum_{\langle ij \rangle \sigma} c_{i\sigma}^{\dag}c_{j\sigma}
 -t^{\prime \prime}\sum_{\langle ij_2 \rangle \sigma}
 c_{i\sigma}^{\dag}c_{j_2\sigma}
 + \sum_{\langle ij \rangle} \left[J \left({\bf S}_i{\bf S}_j-
 {1\over 4}n_in_j\right)+Vn_in_j\right].
 \end{equation}
 $c_{i \sigma}^{\dag}$ is the  creation operator of an electron with
 spin $\sigma$ $(\sigma =\uparrow, \downarrow)$ at site $i$
 of the two-dimensional square lattice. The $c_{i \sigma}^{\dag}$ operators
 act in the Hilbert space with no double electron occupancy.
 The $\langle ij \rangle$ represents nearest neighbor sites,
 and $\langle ij_2 \rangle$ represents next next
 nearest sites. The spin operator is ${\bf S}_i={1\over 2}\sum_{\alpha,\beta}
 c_{i \alpha}^{\dag} {\bf \sigma}_{\alpha \beta} c_{i \beta}$,
 and the number density operator is 
 $n_i=\sum_{\sigma}c_{i \sigma}^{\dag}c_{i \sigma}$.
 In addition to the minimal $t-J$ model (see Ref. \cite{Dag}) we have
 introduced additional next next nearest hopping $t^{\prime \prime}$, and 
 Coulomb repulsion $V$ at nearest sites. Note that we do not introduce
 next nearest neighbor hopping $t^{\prime}$ (diagonal) because we do not
 need it for the purposes of this study.

 In the present work we consider the perturbation theory limit 
 $t, t^{\prime \prime} \ll J$. It is well known that the $t-J$ model
 with small hopping has phase separation at doping. The reason for this
 is very simple: attraction between holes at nearest sites due to
 the reduction in number of missing AF links. The value of this attraction
 immediately follows from eq.(\ref{H}): $U\approx J\langle {\bf S}_i{\bf S}_j-
 1/4\rangle \approx -0.58J$. This effect is very simple and is not
 of interest to us. To overcome it we choose a Coulomb repulsion $V=0.58J$.
 So the only purpose of introducing the Coulomb repulsion in the model
 is to eliminate short range hole-hole attraction.
 
 We consider phase with long range AF order. It is well
 known (see e.g. Ref. \cite{Shr}) that minima of the single hole dispersion 
 are at the faces of the magnetic Brillouin zone $(\pm \pi/2,\pm \pi/2)$, and 
 the dispersion is (we take energy at the minimum as a reference point)
 \begin{eqnarray}
 \label{e}
 &&\epsilon_k= \beta_1\gamma_{\bf k}^2+\beta_2(\gamma^-_{\bf k})^2,\\
 &&\beta_1=3.26t^2/J+8t^{\prime \prime}, \ \ \ 
\beta_2=1.21t^2/J+8t^{\prime \prime}.
 \nonumber
 \end{eqnarray}
 $\gamma_{\bf k}={1\over 2}(\cos k_x + \cos k_y)$,
 $\gamma_{\bf k}^-={1\over 2}(\cos k_x - \cos k_y)$.
Calculation of the $t^{\prime \prime}$ contribution
 in the dispersion (\ref{e}) is straightforward because it is hopping within
 the same magnetic sublattice. Calculation of the $t$-contribution is more 
 complicated, and we use the results of a series expansion\cite{Zheng}.
 Note that a selfconsistent Born approximation \cite{Kane} gives
 substantially different coefficients ($\beta_1=1.86t^2/J$, 
 $\beta_2=0.26t^2/J$). The reason is that there are several 
 important diagrams at small $t$ which are missed in this approximation.
 We will consider the case of very small doping, $\delta \ll 1$, with respect to
 half filling (total filling is $1-\delta$). In this case
 all holes are concentrated in small pockets around the points
 ${\bf k_0}=(\pm \pi/2,\pm \pi/2)$. Single hole dispersion (\ref{e}) can be
 expanded near each of these points
 $\epsilon_k= \beta_1 p_1^2/2+\beta_2 p_2^2/2$, where ${\bf p}$ is deviation
 from the center of the pocket: ${\bf p}={\bf k-k_0}$, $p_1$ is orthogonal to 
 the face of the magnetic Brillouin zone, and $p_2$ is parallel to the face.
 The Fermi energy and Fermi momentum for the holes equal
 $ \epsilon_F \approx \frac{1}{2} \pi (\beta_1 \beta_2)^{1/2} \delta$, \ 
 $p_F \approx \sqrt{p_{1F}p_{2F}} \approx (\pi \delta)^{1/2}$.

 Spin-wave excitations on an AF background are usual spin waves
 with dispersion $\omega_{\bf q}=2J\sqrt{1-\gamma_{\bf q}^2}
 \approx \sqrt{2}Jq$,  at  $q << 1$, see 
 Ref.  \cite{Man} for review.The hole-spin-wave interaction is well known
 (see, e.g. Ref.\cite{Kane})
 \begin{eqnarray}
 \label{hsw}
 && H_{h,sw} = \sum_{\bf k,q} g_{\bf k,q}
     \biggl(
   h_{{\bf k}+{\bf q}\downarrow}^{\dag} h_{{\bf k}\uparrow} \alpha_{\bf q}
   + h_{{\bf k}+{\bf q}\uparrow}^{\dag} h_{{\bf k}\downarrow} \beta_{\bf q}
   + \mbox{H.c.}   \biggr),\\
 && g_{\bf k,q} = 4t\sqrt{2}
  (\gamma_{\bf k} U_{\bf q} + \gamma_{{\bf k}+{\bf q}} V_{\bf q}),\nonumber
 \end{eqnarray}
 where $h_{{\bf k}\sigma}^{\dag}=c_{{\bf k},-\sigma}$ is the hole
 creation operator, $\alpha_{\bf q}^{\dag}$ and $\beta_{\bf q}^{\dag}$
 are the spin wave creation operators for $S_z=\mp 1$, and
 $U_{\bf q}=\sqrt{{J\over{\omega_{\bf q}}}+{1\over 2}}$ and
 $V_{\bf q}=-sign(\gamma_{\bf q})\sqrt{{J\over{\omega_{\bf q}}}-{1\over 2}}$
 are parameters of the Bogoliubov transformation diagonalizing spin-wave
 Hamiltonian, see Ref.\cite{Man}. 
 To describe renormalization of the spin wave under doping, it is convenient
 to introduce the set of Green's functions \cite{Igar}
 \begin{eqnarray}
 \label{GFs}
 D_{\alpha \alpha}(t,{\bf q})&=&
 -i\langle T[\alpha_{\bf q}(t)\alpha_{\bf q}^{\dag}(0)]\rangle, \\
 D_{\alpha \beta}(t,{\bf q})&=&
 -i\langle T[\alpha_{\bf q}(t)\beta_{\bf -q}(0)]\rangle, \nonumber \\
 D_{\beta \alpha}(t,{\bf q})&=&
 -i\langle T[\beta_{\bf -q}^{\dag}(t)\alpha_{\bf q}^{\dag}(0)]\rangle,
 \nonumber \\
 D_{\beta \beta}(t,{\bf q})&=&
 -i\langle T[\beta_{\bf -q}^{\dag}(t)\beta_{\bf -q}(0)]\rangle.
 \nonumber
 \end{eqnarray}
 In the present work we consider only the long-range dynamics:
 $q \sim k \sim p_F \ll 1$. In this limit all possible polarization
 operators coincide \cite{Sus}
 $P_{\alpha \alpha}(\omega,{\bf q})=P_{\alpha \beta}(\omega,{\bf q})=
 P_{\beta \alpha}(\omega,{\bf q})=P_{\beta \beta}(\omega,{\bf q})=
 \Pi(\omega,{\bf q})$.
 For stability of the system the condition (Stoner criterion)
 \begin{equation}
 \label{stab}
 \omega_q + 2 \Pi(0,{\bf q}) > 0
 \end{equation}
 must be fulfilled \cite{Sus1}. Otherwise the Green's functions (\ref{GFs}) 
 would possess poles with imaginary $\omega$. 
 Considering holes as a normal Fermi liquid one can easily
 calculate the polarization operator at $q \ll p_F$: 
 $\Pi(0,{\bf q})\approx -4t^2\sqrt{2} q/\pi\sqrt{\beta_1 \beta_2}$,
 Ref. \cite{Sus1}. Relatively weak pairing, which we consider below, does not 
 influence this result. Then the condition of stability can be rewritten as
 \begin{equation}
 \label{stab1}
 \sqrt{\beta_1 \beta_2}  > {{8t^2}\over{\pi J}}.
 \end{equation}
 The dispersion (\ref{e}) at $t^{\prime\prime}=0$ violates this
 condition. This is a well known statement about the instability of
 long-range AF order under doping in a pure $t-J$ model.
 To provide stability we have to choose 
 $t^{\prime\prime} > t^{\prime\prime}_c = 0.0635 t^2/J$.
 We want to have the system in the phase with long range AF order, but
 close to the point of instability, and therefore we will consider
 $t^{\prime\prime}$ only slightly above the critical value. As a measure
 of this closeness it is convenient to introduce the parameter $\eta$
 \begin{equation}
 \label{eta}
 \eta^2=1-{{8t^2}\over{\pi J\sqrt{\beta_1\beta_2}}}\approx 
 3.4{{J(t^{\prime\prime}-t^{\prime\prime}_c)}\over{t^2}} \ll 1.
 \end{equation}
 The criterion (\ref{stab}) is proportional to this parameter.

 Superconducting pairing mediated by the spin-wave exchange 
 can be found analytically \cite{Flam}. Only the pairing within one
 pocket is important. There is no solution for the gap without nodes
 (s-wave), and there is an infinite number of solutions with the nodes.
 The pairing is maximum for the case of a single node line in the pocket,
 and the gap at the Fermi surface ($\epsilon_F= \beta_1 p_1^2/2+\beta_2 p_2^2/2$)
 is of the form
 \begin{eqnarray}
 \label{gap}
 \Delta(\phi)&=&\Delta_0 \sin \phi,\\
 \Delta_0&=& C \epsilon_F e^{-1/g},\nonumber
 \end{eqnarray}
 where
 \begin{eqnarray}
 \label{fi}
  \sin \phi&=&{{\sqrt{\beta_2}p_2}\over{\sqrt{\beta_1 p_1^2+\beta_2 p_2^2}}},\\
 g&=&{{8t^2}\over{\pi J\beta_2(\sqrt{\beta_1/\beta_2}+1)^2}},\nonumber
 \end{eqnarray}
 and $C\sim 1$ is some constant.
 Energy spectrum and Bogoliubov parameters are given by the usual BCS
 formulas
 \begin{eqnarray}
 \label{BCS}
 E_{\bf k}&=&\sqrt{(\epsilon_k-\epsilon_F)^2+\Delta_{\bf k}^2},\\
 u_{\bf k}^2, v_{\bf k}^2
 &=&{1\over 2}\left(1\pm{{\epsilon_{\bf k}-\epsilon_F}\over
 {E_{\bf k}}}\right)\nonumber
 \end{eqnarray}
 Substituting values $\beta_1=3.77t^2/J$ and $\beta_2=1.78 t^2/J$ corresponding
 to the critical point, we find $g \approx 0.25$, and hence
 \begin{equation}
 \label{gap1}
 \Delta_0 \approx \epsilon_F e^{-4} \approx 0.07 {{t^2 \delta}\over{J}}.
 \end{equation}
 The eqs.(\ref{gap}),(\ref{fi}), and (\ref{gap1}) 
 describe pairing within a single pocket. There are effectively two pockets 
 in the magnetic
 Brillouin zone. Taking symmetric and antisymmetric combinations
 between pockets we get d- and g-wave pairings respectively
 (see Ref.\cite{Flam} for details). The gaps, as well as the critical
 temperatures for d- and g-wave, are practically the same. Note that we discuss  the situation
 without short range hole-hole interaction, which has been eliminated by
 adjusting the nearest sites hole-hole Coulomb repulsion ($V\approx0.58J$, 
 see above). If one destroys this adjustment, the degeneracy between d- and 
 g-waves is also destroyed. The crucial point is that the g-wave is not sensitive at all 
 to the  interaction at nearest sites, whereas the d-wave is very 
 sensitive. Therefore at $V < 0.58 J$ the d-wave pairing is enhanced, while
 on the contrary, at larger Coulomb repulsion $V > 0.58 J$ the d-wave is suppressed and
 disappears very fast.
 
 Now we can discuss the spectrum of the spin-triplet excitations. The Spin wave
 polarization operator due to the mobile holes is of the form
 (see e.g. Ref.\cite{Sus})
 \begin{equation}
 \label{pol}
 \Pi(\omega,{\bf q})=\sum_{\bf k,k_0}g_{\bf k_0 q}^2{{2(E_{\bf k}+E_{\bf k+q})}
 \over{\omega^2-(E_{\bf k}+E_{\bf k+q})^2}}
 \left(u_{\bf k}^2v_{\bf k+q}^2+u_{\bf k}v_{\bf k}u_{\bf k+q}v_{\bf k+q}
 \right).
 \end{equation}
 It accounts for both normal and anomalous fermionic Green's
 functions. Eq. (\ref{pol}) includes summation over
 pockets ${\bf k_0}=(\pi/2,\pm\pi/2)$. In these pockets
 the vertex is $g_{\bf k_0,q} \approx 2^{5/4}t(q_x \pm q_y)/\sqrt{q}$.
 Let us consider the case of very small momenta and frequencies:
 $v_Fq < \Delta_0$, and $\omega < \Delta_0$.
 In this limit one can put $q=0$ in eq. (\ref{pol}) everywhere except at the
 vertex and therefore the polarization operator can be evaluated
 analytically
 \begin{equation}
 \label{pol1}
 \Pi(\omega,{\bf q})=-{{4t^2\omega_{\bf q}}\over{\pi J\sqrt{\beta_1\beta_2}}}
 \left(1+i{{\pi\omega}\over{8\Delta_0}}\right)
 \end{equation}
 Note that the the imaginary part is nonzero even at $\omega < 2\Delta_0$
 because the gap (\ref{gap}) has a line of nodes.
 Any of the Green's functions (\ref{GFs}) have a denominator
 $\omega^2-\omega_{\bf q}^2-2\omega_{\bf q}\Pi(\omega,{\bf q})$,
 see e.g. Refs. \cite{Igar,Sus}. The zero of this denominator gives the energy
 and width of the spin-triplet collective excitation. Using eqs.(\ref{pol1})
 and (\ref{eta}) we find
 \begin{eqnarray}
 \label{o}
 o_{\bf q}&=&\eta \omega_{\bf q},\\
 \Gamma_{\bf q}&=&{{\pi\omega_{\bf q}}\over{8\Delta_0}}o_{\bf q}.\nonumber 
 \end{eqnarray}
 In essence this is the renormalized spin-wave, but its energy is much smaller
 than the energy of the bare spin-wave, $o_{\bf q}/\omega_{\bf q}=\eta \ll 1$.
 This collective excitation exists as a narrow peak only at very small $q$
 when $\pi \omega_{\bf q}/8\Delta_0 <1$. At higher $q$ the width is larger
 than its frequency because of decay to particle hole excitations.
 
 The last question we want to discuss is how close to the point of AF instability
 can we approach using the present description, or in other words how small
 can $\eta$ be? The superconducting pairing itself is not
 sensitive to the AF instability and it survives at the transition to the
 disordered phase. The reason for this is clear: even in the disordered phase the
 magnetic correlation length $\xi_M$ is much larger than the typical distances
 $r \sim 1/p_F \sim 1/(\pi \delta)^{1/2}$ which are important for pairing.
 However the spin-triplet spectrum (\ref{o}) is valid only if $\eta$ is not too
 small. The sublattice staggered magnetization $m$ is renormalized because
 of spin-wave spectrum renormalization, see Ref. \cite{Sus}
 \begin{equation}
 \label{m}
 \delta m= -2J\int\left({1\over{o_{\bf q}}}-{1\over{\omega_{\bf q}}}\right)
 {{d^2q}\over{(2\pi)^2}}.
 \end{equation}
 The integral converges at the momenta $q$ where the collective excitation
 exists as a narrow mode: $q < \Delta_0/J$. Our consideration is valid if
 the renormalization of staggered magnetization is small: $\delta m \ll 0.5$.
 Together with eq. (\ref{m}) this gives the limit for validity of our approach.
 \begin{equation}
 \label{eta1}
 \eta \gg \Delta_0/J
 \end{equation}
 At $\Delta_0 >\eta/J > 0$ equation (\ref{o}) for the spectrum is not valid, 
but  nevertheless there is some gapless spin-triplet excitation. The point 
 $\eta=0$ indicates transition to the spin liquid phase and a spin-wave gap 
 is opened here.
 
 In conclusion we have considered a close to half filling 
 $t-J-J^{\prime \prime}-V$ model with Coulomb repulsion $V=0.58J$ adjusted to 
 eliminate short range hole-hole interaction. We restrict our consideration by
 the case of small $t$, $t \ll J$, and small doping $\delta \ll 1$.
 It is demonstrated that at $t^{\prime\prime} > 0.0635t^2/J$ the Neel order is 
 preserved under the doping, and at $t^{\prime\prime} < 0.0635t^2/J$ the 
order is
 destroyed and the system undergoes transition to the spin liquid phase. In 
 both phases there is  d-g-wave superconducting pairing mediated by
 spin-wave excitations.
 In the Neel state we found, analytically, collective spin triplet excitation.
 It exists as a narrow mode only at very small momenta and its energy is
 substantially below the energy of the bare spin wave.


 We wish to thank M. Kuchiev for stimulating discussions.

 \end{document}